\documentclass[amsmath, amssymb,10pt,aps,prb,twocolumn,notitlepage,showpacs,superscriptaddress]{revtex4-2}
\usepackage{amsmath}
\usepackage{graphicx}
\usepackage{calc}
\usepackage{braket}
\usepackage{ulem}   
\usepackage{slashed}
\usepackage{wasysym}
\usepackage{siunitx}
\usepackage{dsfont}
\usepackage{amsthm,amsmath,amsfonts,amssymb,verbatim,color}
\usepackage{graphicx}
\usepackage{subfigure}
\usepackage{bm}
\usepackage{epsfig,slashed}
\usepackage[T1]{fontenc}
\usepackage[colorlinks=true,citecolor=blue,linkcolor=blue,urlcolor=blue]{hyperref}
\usepackage[version=4]{mhchem}
\begin{document}
\title{Observation of the superconducting proximity effect from surface states in SmB$_6$/YB$_6$ thin film heterostructures via terahertz spectroscopy}

\author{Jonathan Stensberg}
\affiliation{Department of Physics and Astronomy, University of Pennsylvania, Philadelphia, Pennsylvania 19104, U.S.A}
\author{Xingyue Han}
\affiliation{Department of Physics and Astronomy, University of Pennsylvania, Philadelphia, Pennsylvania 19104, U.S.A}
\author{Seunghun Lee}
\affiliation{Department of Materials Science and Engineering, University of Maryland, College Park, MD 20742, USA.}
\affiliation{Department of Physics, Pukyong National University, Busan 48513, Republic of Korea}

\author{Stephen A. McGill}
\affiliation{National High Magnetic Field Laboratory, FSU, Tallahassee, Florida 32310, USA}
\author{Johnpierre Paglione}
\affiliation{Maryland Quantum Materials Center, University of Maryland, College Park, MD 20742, USA.}
\author{Ichiro Takeuchi}
\affiliation{Department of Materials Science and Engineering, University of Maryland, College Park, MD 20742, USA.}
\affiliation{Maryland Quantum Materials Center, University of Maryland, College Park, MD 20742, USA.}
\author{Charles L. Kane}
\affiliation{Department of Physics and Astronomy, University of Pennsylvania, Philadelphia, Pennsylvania 19104, U.S.A}
\author{Liang Wu}
\email{liangwu@sas.upenn.edu}
\affiliation{Department of Physics and Astronomy, University of Pennsylvania, Philadelphia, Pennsylvania 19104, U.S.A}


\date{\today}

\begin{abstract}

The AC conduction of epitaxially-grown SmB$_6$ thin films and superconducting heterostructures of SmB$_6$/YB$_6$ are investigated via time domain terahertz spectroscopy. A two-channel model of thickness-dependent bulk states and thickness-independent surface states accurately describes the measured conductance of bare SmB$_6$ thin films, demonstrating the presence of surface states in SmB$_6$. While the observed reductions in the simultaneously-measured superconducting gap, transition temperature, and  superfluid density of SmB$_6$/YB$_6$ heterostructures relative to bare YB$_6$ indicate the penetration of proximity-induced superconductivity into the SmB$_6$ overlayer; the corresponding SmB$_6$-thickness independence between different heterostructures indicates that the induced superconductivity is predominantly confined to the interface surface state of the SmB$_6$. This study demonstrates the ability of terahertz spectroscopy to probe proximity-induced superconductivity at  an interface buried within a heterostructure, and our results show that SmB$_6$ behaves as a predominantly insulating bulk surrounded by conducting surface states in both the normal and induced-superconducting states in both terahertz and DC responses, which is consistent with the topological Kondo insulator picture. 
\end{abstract}

\pacs{}
\maketitle

%
%
\textbf{Introduction} 

SmB$_6$ has long been identified as a mixed-valence Kondo insulator with an anomalous low-temperature resistance plateau that eluded explanation \cite{Menth1969,Nickerson1971,Mott1974,Varma1976}. Following the discovery of topological insulators \cite{Kane2005a,Kane2005b,Fu2007,Hasan2010}, it was proposed that this anomalous resistance plateau is due to topologically protected surface states, making SmB$_6$ the first topological Kondo insultator (TKI) \cite{Dzero2010,Takimoto2011,Dzero2012,Alexandrov2013}. Following this prediction, a flurry of experiments have investigated the basic features of such a TKI \cite{Zhang2013,Kim2013,Wolgast2013,Kim2014,Phelan2014,Syers2015,Chen2015,Nakajima2015,Lee2016,Stern2017, Neupane2013, Jiang2013,Frantzeskakis2013,XuPRB2014,Pirie2019,Ohtsubo2019,Xu2014}, yet despite the evidence in support of the TKI prediction, controversy has continued to surround SmB$_6$ \cite{Erten2016,Li2014,Xiang2017,Tan2015,Hartstein2018,Hartstein2020,Chowdhury2018,Xu2016,Erten2017,Knolle2015,Hlawenka2018,Eo2020,Matt2020}. Much recent work has therefore been dedicated to understanding experimental discrepancies and harmonizing results. Numerous studies have now highlighted common extrinsic issues with studies of bulk crystals, including subsurface cracks in polished bulk samples \cite{Eo2020}, aluminum inclusions in crystals grown by the aluminum flux  method \cite{Thomas2019}, residual bulk conduction attributed to one dimensional crystalline dislocations \cite{Eo2018,Eo2019,Eo2020,Eo2021}, and localized metallic islands around sample impurities \cite{Jiao2018,Souza2020}. Furthermore, previous 
terahertz studies \cite{Laurita2016,Laurita2018} of SmB$_6$ starkly diverged from DC transport by finding an anomalously large AC conductivity without evidence for surface states. These results created a confused picture of SmB$_6$ with radically different AC and DC behaviors that has been frequently invoked by both experimental and theoretical efforts  \cite{Hartstein2018,Eo2019,Souza2020,Chowdhury2018}. However, these terahertz studies were performed using polished bulk crystals that may suffer from the confounding effects mentioned previously and may therefore be reporting extrinsic behaviors.

\begin{figure*}
    \centering
    \includegraphics[width=\textwidth]{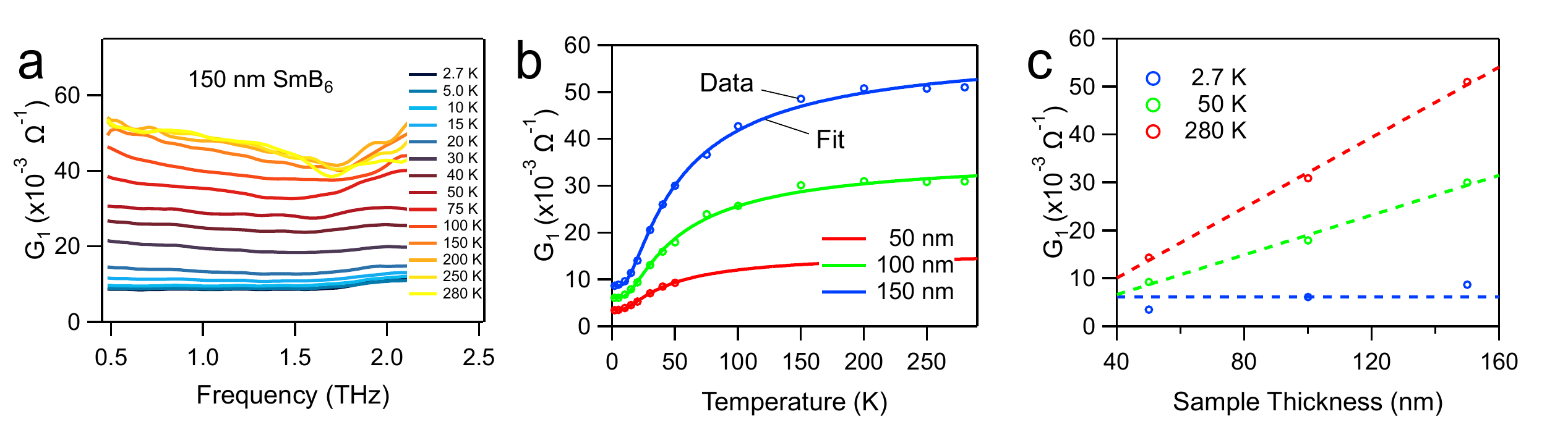}
    \caption{\textbf{a.} The real part of the complex conductance of a 150 nm expitaxial thin film of SmB$_6$ from 2.7 K to 280 K. Complete data for all three samples is provided in Fig. S2 of the SI. \textbf{b.} The average conductance over 0.5-1.0 THz for each SmB$_6$ sample fitted by the two channel conductance model across the temperature range.  \textbf{c.} Thickness dependence of the average conductance at various temperatures for each SmB$_6$ sample. The thickness dependence of all temperatures is provided in Fig. S4 of the SI.}
    \label{fig:Fig1}
\end{figure*}

 Whereas most experiments on SmB$_6$ have employed bulk crystals, it has recently become possible to grow high-quality epitaxial thin films of SmB$_6$ via sputtering \cite{Lee2016,Lee2019,Bae2019}, thereby avoiding the myriad extrinsic concerns with bulk crystals and circumventing issues \cite{Phelan2016} in comparing previous results achieved via the different bulk crystal growth methods. By forming thin film heterostuctures of SmB$_6$ with the isostructural BCS superconductor YB$_6$, perfect Andreev reflection has been observed at the surface of sufficiently-thin SmB$_6$ overlayers via point contact Andreev reflection (PCAR) spectroscopy \cite{Lee2019}. These results indicate the presence of metallic surface states susceptible to the superconducting proximity effect in these epitaxially grown SmB$_6$ samples and, moreover, indicate that these surface states are indeed topologically protected in accord with the TKI prediction.  Such heterostuctures are predicted to host topological superconducting states at the buried interface \cite{Hasan2010,Fu2008} and could be engineered to generate and manipulate Majorana modes to perform topological quantum computations \cite{Hasan2010,Fu2008}. However, such buried interface states are not accessible by standard surface probes such as angle resolved photoemission spectroscopy (ARPES), scanning tunneling spectroscopy/microscopy (STS/M), or  PCAR spectroscopy.

Here, we perform time-domain terahertz spectroscopy (TDTS) on epitaxially grown thin films of SmB$_6$ and SmB$_6$/YB$_6$ heterostructures. We find  an AC conductivity in harmony with DC transport results, demonstrating strong evidence for the presence of surface states in SmB$_6$ at low temperatures and the confinement of the superconducting proximity effect to the surface state at the interface of the SmB$_6$/YB$_6$ heterostructures. Altogether, we establish a straightforward and unified understanding of the intrinsic low temperature conductance of SmB$_6$: in both the normal and induced-superconducting states, SmB$_6$ behaves as a predominantly insulating bulk surrounded by conducting surface states in both AC and DC, as expected under the TKI prediction.

\textbf{Results and Discussion}

 Thin film samples of SmB$_6$ are grown epitaxially on Si (001) substrates via sputtering \cite{Lee2019}. In order to form a minimal-barrier interface with SmB$_6$, the isostructural BCS superconductor YB$_6$ is selected for the proximity effect heterostructures. As the superconducting transition temperature $T_C$ of YB$_6$ is maximized in the case of mild boron deficiency \cite{Lee2019}, 100 nm layers of YB$_{5.6}$ are grown on Si (001) substrates via sputtering, which for convenience will be referred to as YB$_6$ throughout. Heterostructures of SmB$_6$/YB$_6$ are fabricated by growing a 20 nm or 100 nm SmB$_6$ overlayer sequentially atop 100 nm YB$_6$ samples \textit{in situ} without breaking vacuum in the sputtering chamber \cite{Lee2019}.

Typical TDTS \cite{NussOrenstein1998} data for the real conductance $G_1$ is shown for the 150 nm SmB$_6$ sample in Fig. \ref{fig:Fig1}.a (Raw TDTS time trace data is provided in Fig. S1 of the supplementary information (SI)). There are no pronounced spectral features across the reliable frequency range of $\sim$0.5-2.3 THz, though there is a mild Drude-like conductivity that decreases in prominence at lower temperatures. Notably, the conductance of the sample plateaus below 5 K across the entire spectral range. In order to compare the conductance between samples, the average of the spectrum is taken from 0.5 THz to 1.0 THz and shown in Fig. \ref{fig:Fig1}.c for select temperatures (See Fig. S4.a of the SI for all temperatures.). At both 50 K and 280 K, the conductance increases linearly with sample thickness, consistent with bulk-dominated behavior, whereas the conductance is nearly independent of sample thickness at 2.7 K, consistent with surface-dominated behavior. The small amount of thickness dependence that remains at low temperature may be due in part to the limited number of samples available for study, but may also result from a small residual bulk conductivity.

\begin{figure*}
    \centering
    \includegraphics[width=\textwidth]{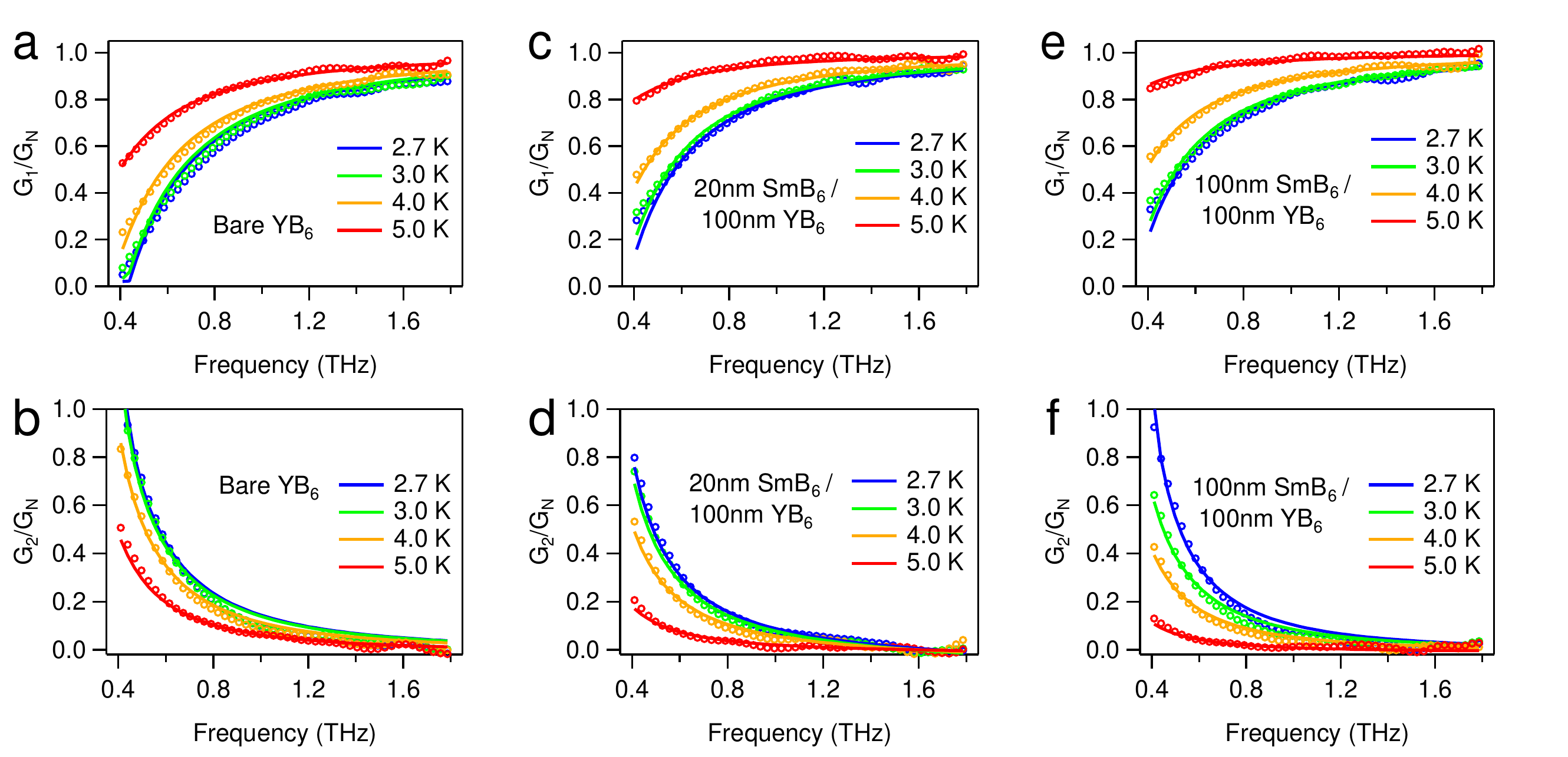}
    \caption{Normalized real and imaginary parts, respectively, of the complex conductance in the superconducting state for bare 100 nm YB$_6$ (\textbf{a,b}), 20 nm SmB$_6$ / 100 nm YB$_6$ (\textbf{c,d}), and 100 nm SmB$_6$ / 100 nm YB$_6$ (\textbf{c,d}). The real and imaginary parts of the data for each sample, given by the circles, are simultaneously fit to produce the solid lines. Unnormalized data for all heterostructures is provided in Fig. S3 of the SI.}
    \label{fig:Fig2}
\end{figure*}

To assess the conductance across the temperature range and available sample thicknesses, we apply a two-channel model of the total conductance $G_{tot}$ \cite{Lee2016,Wolgast2013}. One channel scales with thickness and is exponentially activated as a function of temperature, consistent with a bulk conductance $G_{bulk}$. The second is a temperature- and thickness-independent channel consistent with a surface conductance $G_{surf}$ resulting from both the upper and lower surface states. The two-channel model is thus given by
\begin{equation}\label{EqnTwoChannel1}
    G_{tot}(T) = G_{surf} + G_{bulk}(T)
\end{equation}
\begin{equation}\label{EqnTwoChannel2}
    G_{surf} = G_{LT}
\end{equation}
\begin{equation}\label{EqnTwoChannel3}
    G_{bulk}(T) = \sigma_{bulk,HT}t_{bulk}\rm{exp}(\frac{E_a}{k_B T_{HT}}-\frac{E_a}{k_B T})
\end{equation}
where $G_{LT}$ is the conductance at low temperature, $\sigma_{bulk,HT}$ is the bulk conductivity at high temperature, $t_{bulk}$ is the thickness of the bulk conductance channel, $E_a$ is the characteristic activation energy of the bulk channel, $k_B$ is the Boltzmann constant, and $T_{HT}$ is the temperature at which the high temperature conductivity is calculated. Since the measured low temperature conductance is reasonably consistent across the thin films, in contrast to bulk samples where it can vary by orders of magnitude \cite{Eo2020,Eo2021}, Equations \ref{EqnTwoChannel1}-\ref{EqnTwoChannel3} can be fit to the data while extracting the conductance of each channel, the thickness of each channel, and the bulk activation energy. As can be seen in Fig. \ref{fig:Fig1}.b, the two channel conductance model provides a strong fit to the data for the three samples (experimental data above 50 K for the 50 nm sample proved unreliable possibly due to the substrates being from different batches). The average fitted value of the bulk activation energy $E_a$ = 3.8 meV is consistent with the range of results from previous DC transport measurements on bulk SmB$_6$ crystals \cite{Kim2013,Wolgast2013,Syers2015,Lee2016,Stern2017,Eo2018,Eo2019,Jiao2018}. The fitted values for the thickness of the bulk conductance channel increase linearly with sample thickness in a near one-to-one ratio, indicating the change in conductance between samples is overwhelmingly due to the different thickness of the bulk conducting channel. By considering the actual sample thickness $d = t_{bulk} + 2t_{surf}$, the effective thickness of the surface channel $t_{surf}$ is determined to be consistent and non-negligible, with an average value of $t_{surf}$ = 9.1 nm consistent with previous reports \cite{Lee2016,Bae2019}. This provides strong evidence for surface conducting states in bare SmB$_6$ at low temperature and resolves the previous discrepancy between AC and DC conductance in SmB$_6$.

\begin{figure*}
    \centering
    \includegraphics[width=0.66\textwidth]{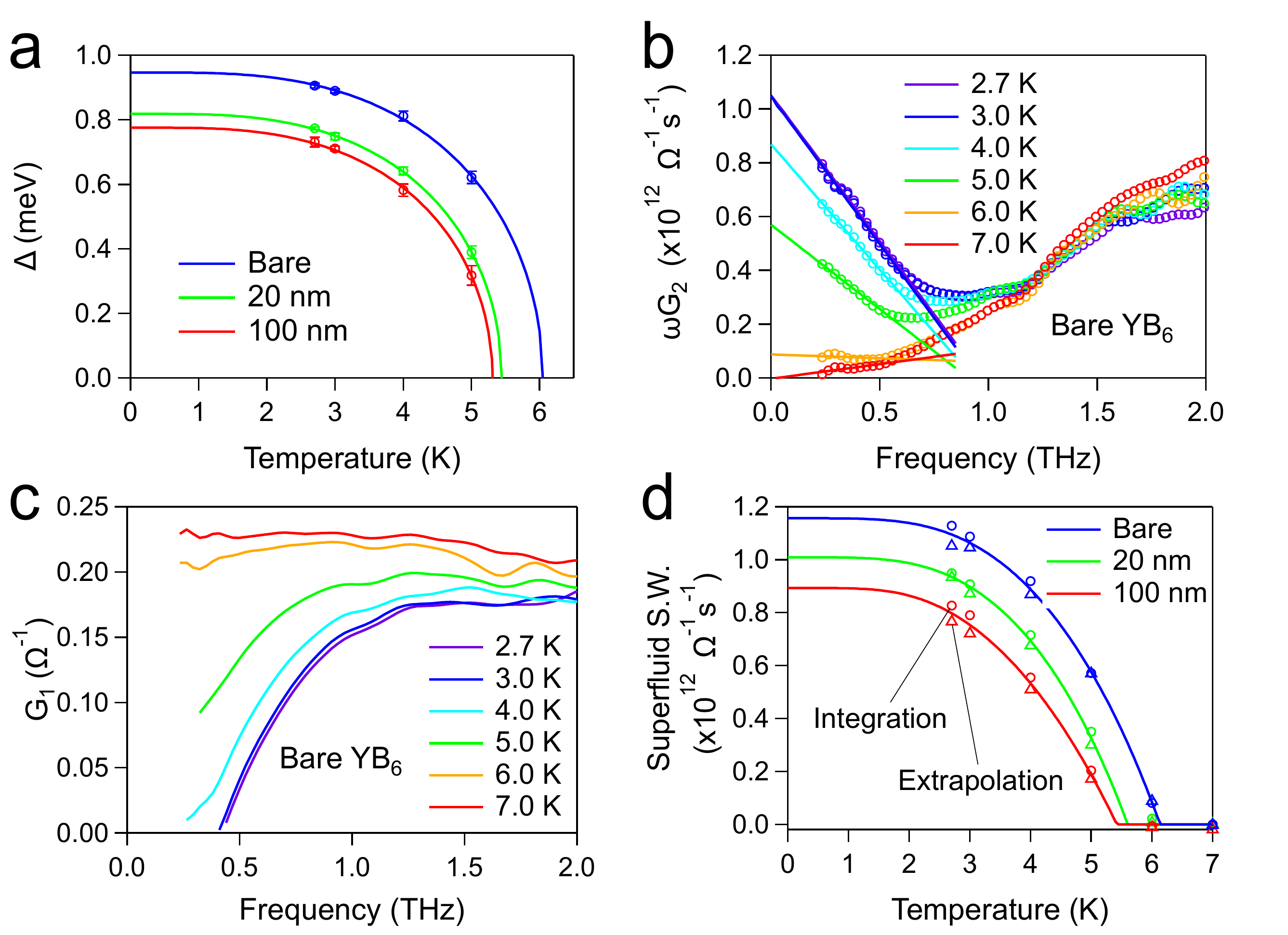}
    \caption{\textbf{a.} BCS fitting of the temperature dependence of the superconducting gap data, $\Delta(T)$, extracted from the Mattis-Bardeen fitting for each superconducting sample. Error bars are determined by the Mattis-Bardeen fitting. \textbf{b.} The linear portion of $\omega G_2$ shown for bare 100 nm YB$_6$ is fitted to permit measurement of the superfluid spectral weight by extrapolation of the fit down to zero frequency. \textbf{c.} The difference in $G_1$ shown for bare 100 nm YB$_6$ is integrated out to 2.0 THz, where the conductance has started to converge, to measure the superfluid spectral weight via integration. \textbf{d.} The superfluid spectral weight determined by the extrapolation and integration methods for each superconducting sample are simultaneously fit by the expected temperature dependence. Error bars for the extrapolation method are smaller than the markers.}
    \label{fig:Fig3}
\end{figure*}

Superconducting heterostructures of SmB$_6$/YB$_6$ are probed via the same TDTS method and compared to a thin film sample of YB$_6$ ($T_C \approx$ 6.1 K) with no overlayer of SmB$_6$.  As all samples consist of 100 nm YB$_6$ and some thickness of SmB$_6$, each heterostructure is referred to by its SmB$_6$ thickness for convenience. Typical data for the bare YB$_6$, the 20 nm heterostructure, and the 100 nm heterostructure are shown in Fig. \ref{fig:Fig2}.a,b; Fig. \ref{fig:Fig2}.c,d; and Fig. \ref{fig:Fig2}.e,f; respectively, where the superconducting low temperature conductance $\Tilde{G} = G_1 + iG_2$  is normalized by the normal state conductance $G_N$ of the sample  above $T_C$ at 10 K. Conductance data of this form is modeled by the Mattis-Bardeen formalism for the optical response of a BCS superconductor in the dirty limit below $T_C$ as the superconducting gap opens \cite{Mattis-Bardeen1958,DresselGruner}. See SI for extended fitting details.

By simultaneously fitting the real and imaginary parts of the normalized conductance for a sample at a given temperature $T$, the superconducting gap $\Delta(T)$ at that temperature can be extracted for a given guess value of $T_C$. By taking an initial estimate of $T_C$ from the disappearance of superconducting behavior in the terahertz spectrum and repeating the simultaneous fitting for each temperature, as shown by the solid lines in Fig. \ref{fig:Fig2}, the temperature evolution of $\Delta(T)$ is extracted. For a BCS superconductor, this temperature evolution is approximated by \cite{Tinkham}
\begin{equation}
    \Delta (T) \approx \Delta_0\rm{tanh}(1.74\sqrt{T_C/T - 1})
\end{equation}
By fitting $\Delta_0$ and $T_C$ to the extracted values of $\Delta(T)$, the guess value of $T_C$ can be updated. Thus by iteratively performing the simultaneous Mattis-Bardeen fitting and BCS gap fitting until convergence, values of $\Delta_0$ and $T_C$ for each sample are extracted from the data. As $\Delta_0$ varies on both sides of the interface of proximity-effect heterostructures \cite{deGennes1964,Clarke1968}, the measured values of $\Delta_0$ are effective averages for the heterostructure.

This iterative method results in a high-quality fit, as shown in Fig. \ref{fig:Fig2} and Fig. \ref{fig:Fig3}.a, with all samples following the BCS behavior.  The clear reduction in both $T_C$ and $\Delta_0$ from the bare YB$_6$ to the heterostructures indicates that superconductivity is being induced in some portion of the SmB$_6$ overlayer via the superconducting proximity effect. For an ordinary metallic overlayer, the reduction in $T_C$ and $\Delta_0$ due to the proximity effect depends strongly on the thickness of the metallic layer for thin films, where the sample thickness is on the order of the normal coherence length, or less \cite{deGennes1964,Clarke1968}. However, the reductions observed in the heterostructures here vary only slightly despite the thickness of the SmB$_6$ considerably spanning the normal coherence length, which was previously determined to be $\sim$50 nm \cite{Bae2019}. The weak SmB$_6$-thickness dependence of the measured $T_C$ and $\Delta_0$ suggests that the effective thickness of the SmB$_6$ that is metallic and susceptible to the proximity effect is largely independent of the actual thickness of the SmB$_6$ overlayer, contrary to the expectation for sample thicknesses on the order of the normal coherence length. This result therefore implies that the dominant contribution to the conductivity is restricted to the surface state at the interface, and that the bulk SmB$_6$ is only weakly conducting at best. Thus the observed weak SmB$_6$-thickness dependence of $T_C$ and $\Delta_0$ in the superconducting heterostructures concords with the model of SmB$_6$ as consisting of metallic surface states surrounding an insulating bulk.

 The measurement of the complex conductance in the superconducting heterostructures also affords a measurement of the superfluid spectral weight, indicating the temperature evolution of the superfluid density in the samples. The superfluid spectral weight can be extracted by two methods, which we will call the extrapolation and integration methods. The extrapolation method makes use of the fact that the superfluid spectral weight is given by \cite{Chauhan2019}
\begin{equation}
    S_{extr}(T) = \lim_{\omega \to 0} \omega G_2^{SC}(\omega,T)
\end{equation}
where $G_2^{SC}$ is the imaginary conductivity in the superconducting state. Extracting values of $S_{extr}$ for each temperature is accomplished by fitting to the linear portion of $\omega G_2(\omega,T)$, as shown in Fig. \ref{fig:Fig3}.b, and extrapolating to zero frequency. The integration method directly calculates the loss of spectral weight when passing below $T_C$ according to \cite{Chauhan2019} 
\begin{equation}
    S_{int}(T) = \int_{0}^{\infty}d\omega(G_1^{N}(\omega) - G_1^{SC}(\omega,T))
\end{equation}
where $G_1^N(\omega)$ and $G_1^{SC}(\omega,T)$ are the real conductivity in the normal state and superconducting states, respectively.  Given the convergence of $G_1$ at high frequency, the upper limit of integration can be reasonably truncated to the limit of reliable data, as shown in Fig. \ref{fig:Fig3}.c, introducing only minor error. Fig. \ref{fig:Fig3}.d shows that while the integration method consistently yields a slightly larger value for the superfluid spectral weight, the two methods show reasonable agreement across the temperature range for each sample. The temperature dependence of the superfluid spectral weight is given by \cite{Chauhan2019,Tinkham}
\begin{equation}
    S(T) = S(0)\frac{\Delta(T)}{\Delta_0}\rm{tanh}(\frac{\Delta(T)}{2k_B T})
\end{equation}

Simultaneous fits of the data for both the extrapolation and integration methods are shown in Fig. \ref{fig:Fig3}.d, showing strong agreement across the temperature range. There is a clear decrease in the superfluid spectral weight between each sample. The decrease from bare YB$_6$ to the heterostructure is expected as a result of the superconducting proximity effect. However, whereas $\Delta_0$ is quite comparable between the heterostructures and shows a difference of just 5 percent, $S(0)$ shows a more significant decrease of 12 percent. The minimal difference in $\Delta_0$ indicates that the proximity effect is predominantly confined to the same volume in both heterostructures, namely the surface states as identified above. The further reduction in $S(0)$ with increased SmB$_6$ thickness, however, may be attributable to very weak conducting states existing in the bulk \cite{Eo2020, Eo2018, Eo2019,Eo2021}. As the superfluid spectral weight is not yet thoroughly explored in the literature, further work is warranted to understand the significance of this behavior.

To summarize,  these results  provide a simple and unified picture in concord with the TKI prediction: SmB$_6$ behaves as a predominantly insulating bulk surrounded by conducting surface states in both the normal and induced-superconducting states in both AC and DC conduction. Experimental explanations and  theoretical speculations that invoked the previous anomalous AC response may need to be reconsidered in light of these findings. While a topologically trivial explanation for this behavior cannot be ruled out by measurements presented here, the previous observation of perfect Andreev reflection \cite{Lee2019} in similar SmB$_6$/YB$_6$ heterostructures supports the topological origin. 

Furthermore, the measurements presented here demonstrate that TDTS can provide an effective probe of superconducting states at the buried interface of these important superconductor-topological insulator heterostructures, providing a powerful new tool for the investigation of engineered topological superconducting systems.  Looking forward, our methods can extend to other topological superconducting heterostuctures with bulk-insulating topological insulators such as Bi$_2$Se$_3$\cite{Koirala2015,Wu2015,Wu2016} and Sb$_2$Te$_3$\cite{Jiang2020} where the proximity effect does not reach the sample surface yet remains active in the buried interface.\\

\textbf{Acknowledgement}

We thank P. Chauhan for helpful discussions. This project is mainly supported by L.W.'s startup package at the University of Pennsylvania. J.S. and X.H. are partially supported by the ARO under the Grants W911NF1910342 and W911NF2020166, and the Gordon and Betty Moore Foundation’s EPiQS Initiative, Grant GBMF9212 to L.W.  The acquisition of the laser for the THz system is supported from a seed grant at the National Science Foundation supported University of Pennsylvania
Materials Research Science and Engineering Center (MRSEC)(DMR-1720530). J.S. is also partially supported by the NSF EAGER grant via the CMMT program (DMR-2132591). C.L.K is supported by a Simons Investigator grant from the Simons Foundation. S.L., J.P., and I.T. are supported by AFOSR FA9550-14-10332. L.W. acknowledges support from the NHMFL Visiting Scientist Program and partial summer support from the NSF EAGER grant.

\bibliography{smb6_bib.bib}
\end{document}


\title{Supplemental Information: Observation of the superconducting proximity effect from surface states in SmB$_6$/YB$_6$ thin film heterostructures via terahertz spectroscopy}

\author{Jonathan Stensberg}
\affiliation{Department of Physics and Astronomy, University of Pennsylvania, Philadelphia, Pennsylvania 19104, U.S.A}
\author{Xingyue Han}
\affiliation{Department of Physics and Astronomy, University of Pennsylvania, Philadelphia, Pennsylvania 19104, U.S.A}
\author{Seunghun Lee}
\affiliation{Department of Materials Science and Engineering, University of Maryland, College Park, MD 20742, USA.}
\affiliation{Department of Physics, Pukyong National University, Busan 48513, Republic of Korea}

\author{Stephen A. McGill}
\affiliation{National High Magnetic Field Laboratory, FSU, Tallahassee, Florida 32310, USA}
\author{Johnpierre Paglione}
\affiliation{Maryland Quantum Materials Center, University of Maryland, College Park, MD 20742, USA.}
\author{Ichiro Takeuchi}
\affiliation{Department of Materials Science and Engineering, University of Maryland, College Park, MD 20742, USA.}
\affiliation{Maryland Quantum Materials Center, University of Maryland, College Park, MD 20742, USA.}
\author{Charles L. Kane}
\affiliation{Department of Physics and Astronomy, University of Pennsylvania, Philadelphia, Pennsylvania 19104, U.S.A}
\author{Liang Wu}
\email{liangwu@sas.upenn.edu}
\affiliation{Department of Physics and Astronomy, University of Pennsylvania, Philadelphia, Pennsylvania 19104, U.S.A}

\date{\today}

\maketitle


\textbf{Time-Domain Terahertz Spectroscopy}

The raw observable in time-domain terahertz spectroscopy (TDTS) is the electric field profile of a terahertz pulse mapped out as a function of time. Here, terahertz pulses are produced by focusing a femtosecond pulsed laser (780 nm, 82 fs pulse duration; 10 mW power; 80 MHz repetition rate) onto a photoconductive antenna (sometimes called an Auston switch). The focused femtosecond pulses generate photoexcited carriers, which are accelerated to produce a brief current in the sub-picosecond scale, when interacting with a periodic external bias field across the antenna. Radiation from this current is coupled out to free space, yielding quasi-single cycle pulses of terahertz radiation. These emitted terahertz pulses are collected via a pair of off-axis parabolic mirrors and focused onto the sample inside of an optical cryostat. Having transmitted through the sample, the terahertz pulses are focused by another pair of off-axis parabolic mirrors onto a second photoconductive antenna. A second branch of the femtosecond laser beam is also focused onto this second photoconductive antenna, and the optical path lengths of the terahertz pulse and femtosecond pulse are matched such that the two pulses arrive at the photoconductive antenna near simultaneously. Photoexcited carriers generated by the femtosecond laser are converted into a current across the antenna by the electric field of the terahertz pulse. As the readout of the resulting current is proportional to the electric field of the terahertz pulse, varying the time delay between the femtosecond pulse and the terahertz pulse allows the electric field profile of the terahertz pulse to be mapped out as a function of time. 

\makeatletter 
\renewcommand{\thefigure}{S\@arabic\c@figure}
\makeatother
 \begin{figure*}
    \centering
    \includegraphics[width=\textwidth]{FigureS1v1.pdf}
    \caption{Raw electric field profile of terahertz pulses in the time domain for the Si substrate with (\textbf{a}) bare SmB$_6$ samples at 2.7 K and (\textbf{b}) SmB$_6$/YB$_6$ heterostructures at 2.7 K. The labels in both subfigures indicate the thickness of the SmB$_6$; all heterostructures are composed of the indicated thickness of SmB$_6$ over 100 nm of YB$_6$. All samples are grown on a Si substrate.}
    \label{fig:FigS1}
\end{figure*}

\begin{figure*}
    \centering
    \includegraphics[width=\textwidth]{FigureS2v2.pdf}
    \caption{Real part of the complex conductance of (\textbf{a}) 50 nm, (\textbf{b}) 100 nm, and (\textbf{c}) 150 nm thick bare SmB6 samples. As mentioned in the main text, the data for the 50 nm sample proved unreliable above 50 K and has been omitted throughout.}
    \label{fig:FigS2}
\end{figure*}

The electric field profile of terahertz pulses through bare SmB$_6$ and SmB$_6$/YB$_6$ heterostructures at low temperature are given in Fig \ref{fig:FigS1}.a,b, respectively, along with the Si substrate.  To calculate the transmission of the sample, the electric field profile through both the sample and the substrate are first Fourier transformed $E(t) \xrightarrow{FFT} E(\omega)$ to yield the electric field spectra $E(\omega)$. The resulting electric field spectra of the sample $E_{sam}(\omega)$ and the substrate $E_{sub}(\omega)$ are then divided: $\Tilde{T}(\omega) = E_{sam}(\omega)/E_{sub}(\omega)$, where $\Tilde{T}(\omega) = T_1(\omega) + iT_2(\omega)$ is the complex transmission of the sample. The transmission is fully complex because the time domain measurement is coherent, keeping track of both the magnitude of the electric field as well as the phase.


For thin films, the complex transmission  $\Tilde{T}(\omega)$ is related to the complex conductance $\Tilde{G}(\omega)$  of the material by the equation

\begin{equation}\tag{S1}\label{EqnS1}
    \Tilde{T}(\omega) = \frac{n+1}{n+1+Z_0\Tilde{G}}e^{i\omega\Delta L(n-1)/c}
\end{equation}

where $n$ is the substrate index of refraction, $Z_0$ is the vacuum impedance, $c$ is the speed of light, and $\Delta L$ is the difference between the thickness of the reference substrate and the thickness of the substrate on which the sample is grown. The exponential term accounts for the change in phase that results from the terahertz pulse having to travel through different lengths of material when there are small differences in substrate thicknesses. The complex conductance $\Tilde{G}(\omega)$ can then be determined by inverting Eqn \ref{EqnS1} to yield

\begin{equation}\tag{S2}\label{EqnS2}
    \Tilde{G}(\omega) = \frac{n+1}{Z_0}(\frac{1}{\Tilde{T}(\omega)}e^{i\omega\Delta L(n-1)/c}-1)
\end{equation}

Thus by plugging in the complex transmission $\Tilde{T}(\omega)$  determined by TDTS for each sample, the complex conductance $\Tilde{G}(\omega)$ of the sample can be determined without reference to the Kramers-Kronig relations. For thin films, it is generally considered $\Tilde{\sigma}$ = $\Tilde{G}/d$, where $d$ is the film thickness and $\Tilde{\sigma}$ is the complex conductivity. While optical measurements are often reported in terms of conductivity, we have elected to report conductance to avoid importing assumptions about the thickness of the surface state $d_s$ into the data and subsequent analysis.

TDTS measurements are performed over a temperature range of 2.7-290 K for all samples, and the real part of the conductance $G_1$ is given in Fig \ref{fig:FigS2} (the data for the 150 nm sample in Fig \ref{fig:FigS2}.c is presented in Fig 1.a of the main text). The data for the 50 nm sample was determined to be unreliable above 50 K. For all TDTS measurements, back-to-back measurements are used to determine the reliable frequency range.

\begin{figure*}
    \centering
    \includegraphics[width=\textwidth]{FigureS3v1.pdf}
    \caption{Real and imaginary parts, respectively, of the complex conductance of (\textbf{a,b}) the bare 100 nm YB$_6$, (\textbf{c,d}) the 20 nm SmB$_6$ / 100 nm YB$_6$ heterostructure, and (\textbf{e,f}) the 100 nm SmB$_6$ / 100 nm YB$_6$ heterostructure.}
    \label{fig:FigS3}
\end{figure*}

\textbf{Mattis-Bardeen Fitting}

The complex conductance of the superconducting SmB$_6$/YB$_6$ heterostructures is determined by the same method. In the case of superconductors, however, key conductance behavior is encoded in both the real part $G_1$ and the imaginary part $G_2$. Consequently, both parts of the complex conductance are shown in Fig \ref{fig:FigS3} for all three heterostucture samples. The data shows typical BCS-like superconducting behavior, where the real conductance $G_1$ is suppressed at low frequency while the imaginary conductance $G_2$ is enhanced, forming a 1/$\omega$-esque frequency dependence. The complex conductance in the superconducting state can be modeled by the Mattis-Bardeen formalism.

The TDTS conductance data below the superconducting transition temperature $T_c$ for the bare YB$_6$ and SmB$_6$/YB$_6$ heterostructure samples are fitted via the Mattis-Bardeen equations, as stated in the main text. The Mattis-Bardeen equations are given as follows

\begin{equation}\label{EqnMB1}\tag{S3}
    \begin{gathered}
    \begin{aligned}
    \frac{G_1(\omega,T)}{G_N} &= \\ \frac{2}{\hbar\omega}&\int_{\Delta}^{\infty}\frac{(f(\epsilon)-f(\epsilon+\hbar\omega))(\epsilon^2+\Delta^2+\hbar\omega\epsilon)}{(\epsilon^2-\Delta^2)^{1/2}((\epsilon+\hbar\omega)^2-\Delta^2)^{1/2}}d\epsilon \\ + \frac{1}{\hbar\omega}&\int_{\Delta-\hbar\omega}^{-\Delta}\frac{(1-2f(\epsilon+\hbar\omega))(\epsilon^2+\Delta^2+\hbar\omega\epsilon)}{(\Delta^2-\epsilon^2)^{1/2}((\epsilon+\hbar\omega)^2-\Delta^2)^{1/2}}d\epsilon
    \end{aligned}
    \end{gathered}
\end{equation}
\begin{equation}\label{EqnMB2}\tag{S4}
    \begin{gathered}
    \begin{aligned}
    \frac{G_2(\omega,T)}{G_N} &=\\ \frac{1}{\hbar\omega}&\int_{\Delta-\hbar\omega,-\Delta}^{\Delta}\frac{(1-2f(\epsilon+\hbar\omega))(\epsilon^2+\Delta^2+\hbar\omega\epsilon)}{(\Delta^2-\epsilon^2)^{1/2}((\epsilon+\hbar\omega)^2-\Delta^2)^{1/2}}d\epsilon
    \end{aligned}
    \end{gathered}
\end{equation}
where $G_1(\omega,T)$ and $G_2(\omega,T)$ are respectively the real and imaginary conductance at temperatures $T < T_C$, $G_N$ is the conductance in the normal state above $T_C$, $f(\epsilon)$ is the Fermi-Dirac function for the energy $\epsilon$ at the given temperature $T$, $\hbar$ is the reduced Planck constant, $\Delta(T)$ is the superconducting gap at the given temperature $T$. The second term in Eqn \ref{EqnMB1} is zero when $\hbar\omega < 2\Delta$, and is as written when $\hbar\omega > 2\Delta$. Likewise, the lower limit of integration in Equation \ref{EqnMB2} is $\Delta-\hbar\omega$ when $\hbar\omega < 2\Delta$, and is $-\Delta$ when $\hbar\omega > 2\Delta$. In the data analysis, $G_N$ is taken at $T = 10$ K, and as can be seen in Fig \ref{fig:FigS3}, the conductance above $T_C$ is not strongly dependent upon temperature.

As described briefly in the main text, the fitting process for a given sample begins by assuming some initial guess value for $T_C$. This is done by identifying when superconducting behavior first appears in the conductance data of Fig \ref{fig:FigS3}. Using this initial guess value of $T_C$, the normalized conductance data (shown in Fig. 2 of the main text) at a given temperature $T < T_C$ is fitted to Eqns \ref{EqnMB1} and \ref{EqnMB2}, which yields a fitted value of the superconducting gap $\Delta(T)$ at that temperature $T$. This fitting is then repeated for each temperature $T$ (using the same guess value of $T_C$), thus yielding a series of $\Delta(T)$ data points over the temperature range $T < T_C$. This series of $\Delta(T)$ is then fitted to the expected temperature dependence of the superconducting gap for a BCS superconductor, given by Eqn 4 of the main text, which yields a fitted value of $T_C$. The fitting procedure then proceeds iteratively by using this fitted value of $T_C$ as the new guess value and repeating the fittings until the value of $T_C$ from successive iterations converges. This iterative process yields high quality fits to the data for all samples, as shown in Fig 2 of the main text, and the resulting fitted data for $\Delta(T)$ is shown in Fig 3.a of the main text.\\

\textbf{Two Channel Conductance Model}

As detailed in the main text, a two-channel model of conductance is applied to the bare SmB$_6$ samples. In brief, one channel models the combined contribution from the top and bottom surface conducting states as a constant, temperature-independent term. The second channel models the contribution from the bulk, being exponentially activated by temperature and linearly proportional to the thickness of the channel. This model is summarized by Eqns 1-3 of the main text. Since the conductance of the bare SmB$_6$ is only mildly frequency-dependent for all three samples, the conductance for the three bare SmB$_6$ samples at each temperature is averaged over 0.5 – 1.0 THz to produce a single data point for each sample at each temperature. To highlight the transition from thickness-dependent conductance at high temperature to nearly thickness-independent conductance at low temperature, these data points where shown in Fig 1.c of the main text, but for clarity, only 2.7 K, 50 K, and 280 K were shown. Fig \ref{fig:FigS4}.a shows these data points for all temperatures, with the convergence to nearly thickness-independent conductance clear at low temperatures. The two-channel conductance model is fit to these data points, and fitted curves are displayed in Fig 1.b of the main text. From the fitting, the thickness of the surface channels and the bulk activation energy are determined for each sample. The values of these parameters are shown in Fig \ref{fig:FigS4}.b for all three samples, along with the average values indicated by the blue and red dashed lines, respectively. The average value of the surface channel thickness is found to be 9.1 nm, and the average value of the activation energy is 3.8 meV, as stated in the main text. Linear extrapolations of the data in \ref{fig:FigS4}.a find a intersection of all the data points around 18 nm and 0.005 ($\Omega^{-1)}$). These values coincide nicely with the fitted values the thickness and conductance of two surface channels, respectively $2t_{surf} = 18.2 $ nm and $G_{surf} = 0.006~ (\Omega^{-1})$, further demonstrating the consistency between the data and the model.

\begin{figure*}
    \centering
    \includegraphics[width=\textwidth]{FigureS4v2.pdf}
    \caption{(\textbf{a}) The real conductance of each bare SmB$_6$ sample at each temperature averaged over the range 0.5 – 1.0 THz (for the 50 nm sample, the data points above 50 K are determined from the fitting of the two-channel conductance model). (\textbf{b}) The fitted values of the surface channel thickness (blue) and activation energy (red) for the three samples. The dashed line indicates the average value for the three sample thicknesses.}
    \label{fig:FigS4}
\end{figure*}